\documentclass[12pt]{iopart}
\usepackage{amssymb}
\usepackage{graphicx}

\begin{document}

\title[The band structure of the whole spectrum of an N-body cold system]
      {The band structure of the whole spectrum of an N-body cold system containing atoms with arbitrary integer spin and dominated by singlet pairing force}

\author{Y Z He$^1$, Y M Liu$^2$, Z B Li$^1$, C G Bao$^{1,}$\footnote{Corresponding author: C G Bao, stsbcg@mail.sysu.edu.cn}}
\address{$^1$School of Physics, Sun Yat-Sen University, Guangzhou, 510275, P. R. China}
\address{$^2$Department of physics, Shaoguan University, Shaoguan, 510205, P. R. China}

\begin{abstract}
The spectra of $N$-boson systems with arbitrary nonzero spin $\mathfrak{f}$ have been studied.
Firstly, only the singlet pairing interaction is considered, a set of eigenstates together with the eigenenergies are analytically obtained.
The completeness of this set is proved.
The analytical expression allows us to see clearly the spin structures of various states different in $N$ and/or $\mathfrak{f}$, and to find out the similarity and relationship lying among them.
Secondly, the effect of other interactions is evaluated via exact numerical calculations on the systems with a smaller $N$.
Some features and notable phenomena that might emerge in high-$\mathfrak{f}$ systems, say, the ground band might have extremely high level density, have been discussed.
\end{abstract}


\submitto{\PS}

\maketitle

\section{Introduction}

The study of many-body cold systems is a hot topic in recent years.
What we pay attention to and think about are the following points:
\begin{itemize}
\item
After the application of optical trapping, the spin degrees of freedom are liberated \cite{ref5,ref6,ref6p,ref7,ref10}.
Furthermore, due to the progress in techniques, very low temperature (say, $T\leq 10^{-11}K$) can be achieved \cite{lt}.
At such a temperature the spatial degrees of freedom can be completely frozen, and the systems dominated by spin degrees of freedom can be experimentally created.
It turns out that the spin dependent forces are very weak (nearly two order weaker than the central force).
Thus, an important feature of these systems is the extreme high sensitivity.
\item
Due to the progress in techniques, the precise manipulation of a few atoms can be realized \cite{few1,few2}.
Thus, instead of a very large particle number $N$, the study of cold systems with a smaller $N$ (say, $N\leq 10$) is meaningful because they might be controlled more precisely.
\item
The role of [0]-pairs, namely, the particles are two-by-two coupled to spin zero, in fermion systems is well known (say, in superconductivity and in nuclear theory) \cite{bcs1,bcs2}.
For spin-$\mathfrak{f}$ boson systems with $\mathfrak{f}\leq 2$, it has been proved that the number of the [0]-pairs is a good quantum number due to the conservation of seniority \cite{ref10,pvi,yka}, and they appear as a basic constituent in spin structures.
The number of them is an important index for classifying the states.
Nonetheless, when $\mathfrak{f}\geq 3$, seniority is in general not conserved (except in some special cases).
Therefore, the role of the [0]-pairs in $\mathfrak{f}\geq 3$ systems remains to be further clarified.
\end{itemize}

While numerous literatures have been dedicated to the spin-1 and spin-2 systems, the study on $\mathfrak{f}\geq 3$ systems is relatively scarce \cite{sp3a,sp3b,sp3c,sp3d,sp3e}.
It turns out that the spin degrees of freedom would increase greatly with $\mathfrak{f}$.
Let $N=10$ for an example.
If $\mathfrak{f}=1$, 2, 3, and 4, the number of spin states with magnetization $M=0$ is 6, 55, 338, and 1514, respectively.
When $N$ becomes larger, the number is terribly large.
Say, when $N=1000$ and $\mathfrak{f}=3$, this number would be $7.7735\times 10^{11}$.
Due to the great increase of the dimension of the spin space, rich physics would be involved in the high-$\mathfrak{f}$ systems.
Thus, the study of them is worthy.

This paper is dedicated to the study of $N$-boson spin-$\mathfrak{f}$ bound cold systems with arbitrary $N$ and $\mathfrak{f}$ (nonzero).
The study aims beyond the ground state but covers the whole spectrum.
The purpose is to clarify the role of singlet pairing in high-$\mathfrak{f}$ systems, and to find out common characters among these systems.

\section{The complete spectra of high-$\mathfrak{f}$ systems under singlet pairing interaction}

We assume that the temperature is so low that all the spatial degrees of freedom are frozen, and all particles fall into the same spatial state $\phi$ which is most favorable to binding.
Due to the freezing, the spin-orbit coupling can be neglected.
Then the high-$\mathfrak{f}$ system is completely governed by the spin dependent Hamiltonian
\begin{eqnarray}
 H_{\mathrm{spin}}
  =  \sum_{i<j}
     V_{ij},\ \ \
 V_{ij}
  =  \sum_{\lambda}
     g_{\lambda}
     P_{\lambda }^{i,j}, \label{h}
\end{eqnarray}
where $g_{\lambda}$ is the weighted strength of the $\lambda$-channel ($\lambda=0,2,\cdots,2\mathfrak{f}$), and the factor $\int\phi^4(r)\mathrm{d}\bi{r}$ has been included.
$P_{\lambda}^{i,j}$ is the projector to the $\lambda$-channel.
Let the Hamiltonian be dominated by singlet pairing interaction, namely, $g_0$ is much more negative than the other strengths.
In this case the effect of the latter is smaller, thus they can be approximately given as being equal to each other.
Since the spin structures will not be changed when the set $\{g_{\lambda}\}$ are shifted as a whole and/or when a new unit is adopted, the case with $g_0<g_2=\cdots=g_{2\mathfrak{f}}$ is equivalent to the case with $g_0=-1$ and $g_2=\cdots=g_{2\mathfrak{f}}=0$.
This simplified Hamiltonian is denoted as $H_{[0]}$.
We are first going to find out all the eigenenergies and eigenstates of $H_{[0]}$ analytically.
Then, when $H_{\mathrm{spin}}$ deviates from $H_{[0]}$, the effect of the deviation is evaluated.

Let $\{\Phi_{N_1,S,l}\}$ be a set of normalized, symmetrized, and orthogonal spin states of a subsystem with $N_1$ particles.
$S$ is the total spin and $l$ is just an index to specify further the states.
Let $\chi$ denote the spin state of a particle, and $(\chi\chi)_0$ denote a singlet pair ([0]-pair).
Let $N=N_1+2J$ and $\mathfrak{P}_N$ denote a summation over the $N!$ permutations of particles.
Then, for the product state $\mathfrak{P}_N\Phi_{N_1,S,l}(\chi \chi)_0^J$, we divide the $N!$ permutations into four parts as
\begin{eqnarray}
 \mathfrak{P}_N
  =  \mathfrak{P}_{N,a}
    +\mathfrak{P}_{N,b}
    +\mathfrak{P}_{N,c}
    +\mathfrak{P}_{N,d},
\end{eqnarray}
where $\mathfrak{P}_{N,x}$ includes those permutations that the particles $i$ and $j$ are both in $\Phi_{N_1,S,l}$ (if $x=a$), $i$ in $\Phi_{N_1,S,l}$ and $j$ in $(\chi\chi)_0^J$ (if $x=b$), $j$ in $\Phi_{N_1,S,l}$ and $i$ in $(\chi\chi)_0^J$ (if $x=c$), and both $i$ and $j$ are in $(\chi\chi)_0^J$ (if $x=d$).
For the case $x=b$, formally we can extract the $i$-th particle from $\Phi_{N_1,S,l}$ by using the fractional parentage coefficients $\beta_{S'}$ as $\Phi_{N_1,S,l}=\sum_{S'}\beta_{S'}(\chi(i)\phi_{S'})_S$.
The details of $\beta_{S'}$ and $\phi_{S'}$ are irrelevant in the follows.
Thus we have
\begin{eqnarray}
 \mathfrak{P}_{N,b}
 [\Phi_{N_1,S,l}(\chi\chi)_0^J]
  =  \mathfrak{P}_{N-2}^{(i,j)}
     2N_1
     J
     [ \sum_{S'}
       \beta_{S'}
       (\chi(i)\phi_{S'})_S
       (\chi(j)\chi)_0
       (\chi\chi)_0^{J-1}],
\end{eqnarray}
where $\mathfrak{P}_{N-2}^{(i,j)}$ denotes a summation over all $(N-2)!$ permutations (particles $i$ and $j$ are excluded).
Making use of the following specific 9-$j$ symbol
\begin{eqnarray}
 \left\{
 \begin{array}{ccc}
   \mathfrak{f} & S' & S \\
   \mathfrak{f} & \mathfrak{f} & 0 \\
   0 & S & S
 \end{array}
 \right\}
  =  \frac{(-1)^{\mathfrak{f}+S'+S}}
          {(2\mathfrak{f}+1)(2S+1)},
\end{eqnarray}
We have
\begin{eqnarray}
 (\chi(i)\phi_{S'})_S
 (\chi(j)\chi)_0
  =  \frac{1}{2\mathfrak{f}+1}
     (\chi(i)\chi(j))_0
     (\chi\phi_{S'})_S
    +Z.
\end{eqnarray}
In $Z$ the particles $i$ and $j$ are not coupled to zero.
When $H_{\mathrm{spin}}=H_{[0]}$, $Z$ has no contribution to energy and therefore can be neglected.
After the neglect,
\begin{eqnarray}
 \mathfrak{P}_{N,b}
 [\Phi_{N_1,S,l}(\chi\chi)_0^{J}]
  =  \frac{2N_1J}{2\mathfrak{f}+1}
     \mathfrak{P}_{N-2}^{(i,j)}
     [\Phi_{N_1,S,l}
      (\chi(i)\chi(j))_0
      (\chi\chi)_0^{J-1}],
\end{eqnarray}
and
\begin{eqnarray}
\fl
 H_{[0]}\mathfrak{P}_{N,b}
 [\Phi_{N_1,S,l}(\chi\chi)_0^J]
  =  \frac{1}{2}
     \sum_{i\neq j}
     V_{ij}
     \mathfrak{P}_{N,b}
     [\Phi_{N_1,S,l}(\chi\chi)_0^J]
  = -\frac{N_1J}{2\mathfrak{f}+1}
     \mathfrak{P}_N
     [\Phi_{N_1,S,l}(\chi\chi)_0^J].
\end{eqnarray}
The above formula remains unchanged when $\mathfrak{P}_{N,b}\rightarrow\mathfrak{P}_{N,c}$.

Similarly, we have
\begin{eqnarray}
 H_{[0]}\mathfrak{P}_{N,d}
 [\Phi_{N_1,S,l}(\chi\chi)_0^J]
  = -\frac{(2J+2\mathfrak{f}-1)J}{2\mathfrak{f}+1}
     \mathfrak{P}_{N}
     [\Phi_{N_1,S,l}(\chi\chi)_0^J].
\end{eqnarray}
Let the set $\{\Phi_{N_1,S,l}\}$ include all those states of the $N_1$-body subsystem full in seniority (i.e., all the $N_1$ particles are unpaired).
Accordingly, $\Phi_{_{N_1,S,l}}$ has no contribution on energy when $H_{\mathrm{spin}}=H_{[0]}$.

Let $\tilde{\mathfrak{P}}_N$ be the operator for normalization and symmetrization, and let
\begin{eqnarray}
 \Psi_{J,S,l}^{[N]}
  \equiv
     \tilde{\mathfrak{P}}_N
     [\Phi_{N_1,S,l}(\chi\chi)_0^J]. \label{wf}
\end{eqnarray}
From the above formulae we arrive at
\begin{eqnarray}
 H_{[0]}\Psi_{J,S,l}^{[N]}
  = -\frac{(2N-2J+2\mathfrak{f}-1)J}
          {2\mathfrak{f}+1}
     \Psi_{J,S,l}^{[N]}
  \equiv
     E_J^{[N]}
     \Psi_{J,S,l}^{[N]}. \label{sch}
\end{eqnarray}
which is just the Schr\"{o}dinger equation.
Thus, a series of eigenenergies and eigenstates of $H_{[0]}$ have been analytically obtained.
The completeness of this set is given below.
Each state is a product of a core (filled with $N_1$ unpaired particles) together with $J$ [0]-pairs.
In particular, when $N$ is fixed, $E_J^{[N]} $ depends only on $J$, thus a group of states with the same $J$ but different in $S$ and $l$ are degenerate.
When $N=2K$ or $2K+1$, $J$ is ranged from 0 to $K$.
Accordingly, the whole spectrum is divided in to $K+1$ bands specified by $J$, the width of each band is zero.
Obviously, when $J$ is larger, more [0]-pairs are contained, therefore the energy $E_J^{[N]}$ is lower.
The lowest band has $J=K$ and contains only one state, namely, the ground state (g.s.) $\Psi_{\mathrm{gs}}^{[N]}=\tilde{\mathfrak{P}}_N[\chi(\chi\chi)_0^J]$ or $\tilde{\mathfrak{P}}_N[(\chi\chi)_0^J]$.
The g.s. energy is $E_{\mathrm{gs}}^{[N]}=-[2(K+\mathfrak{f)}\pm 1]K/(2\mathfrak{f}+1)$, where $+(-)$ is for odd (even) $N$.

\begin{figure}[htb]
 \centering \resizebox{0.95\columnwidth}{!}{\includegraphics{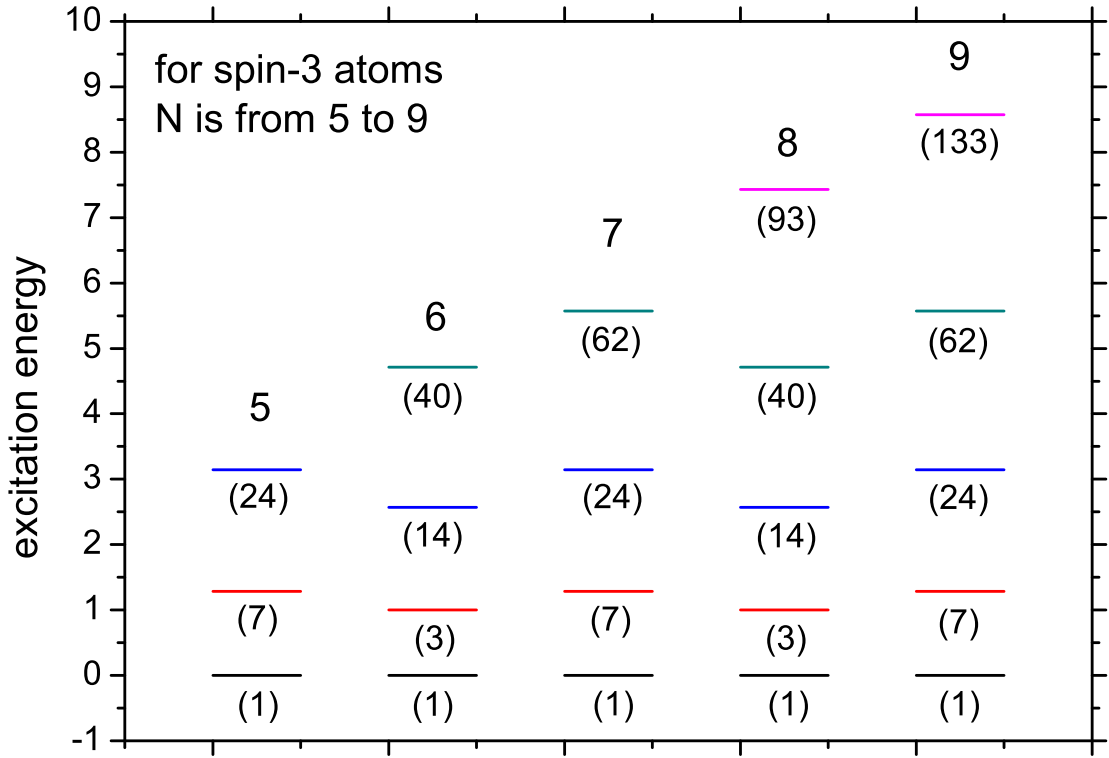} }
 \caption{(color online)
 The complete excitation spectra of spin-3 systems with $N=5$ to 9 ($N$ is marked at the top of each spectrum) and with $H_{\mathrm{spin}}=H_{[0]}$ (i.e., $g_0=-1$, $g_2=g_4=g_6=0$, and the energy unit is $|g_0|$).
 Each spectrum is divided into bands specified by $L=0$ (for the bottom band) to $K$ (for the top band), $K=N/2$ (or $(N-1)/2$), if $N$ is even (or odd).
 The excitation energy $E_L^{[N]\mathrm{ex}}$ is marked by a short horizontal line.
 The number of degenerate states with $M=0$ included in the $L$ band, $\mathfrak{N}_{L,(-1)^N}$, is given below the line.
 Say, the first excited band with $L=1$ and $N=9$ contains seven $M=0$ states.
 }
 \label{fig1}
\end{figure}

\begin{figure}[htb]
 \centering \resizebox{0.95\columnwidth}{!}{\includegraphics{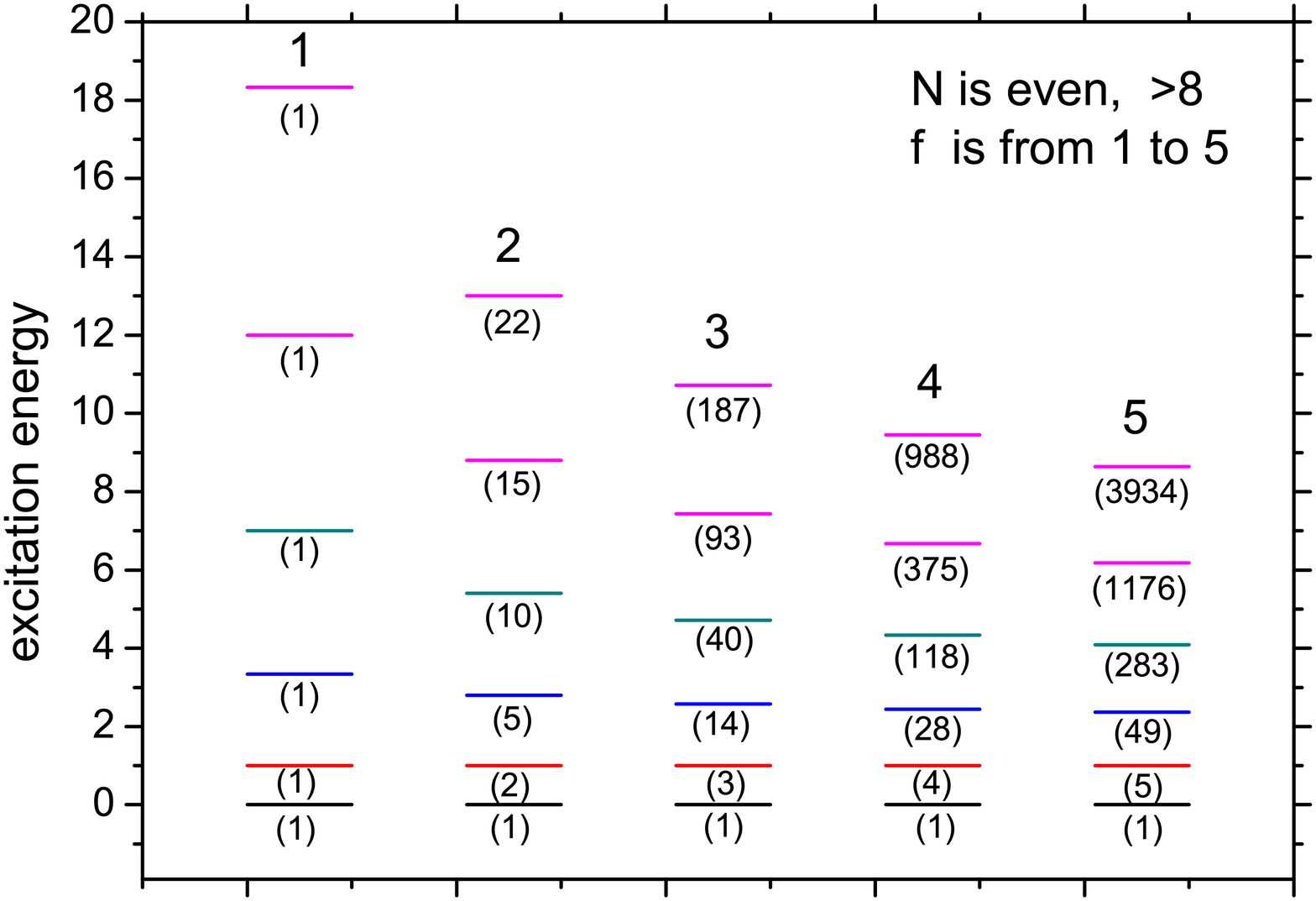} }
 \caption{(color online)
 The lower part of the excitation spectra of spin-$\mathfrak{f}$ systems with $H_{\mathrm{spin}}=H_{[0]}$.
 $\mathfrak{f}$ is from 1 to 5 (marked at the top of each spectrum), $N\geq 10$ (even), and $\mathfrak{N}_{L,(-1)^N}$, is given below the line of each level.
 Note that the spectrum depends on $(-1)^N$ but not on $N$.
 It implies that these spectra hold disregarding how large $N$ is.
 Nonetheless, when $N\leq 8$, some upper bands would disappear.
 }
 \label{fig2}
\end{figure}

Let $L\equiv K-J$.
While $J$ denotes the number of [0]-pairs, $L$ is related to the seniority ($L=N_1/2$ or $(N_1-1)/2$ if $N$ is even or odd).
It denotes also the bands in the order of uprising energy, say, $L=0$ ($K$) is for the bottom (top) band.
The excitation energy
\begin{eqnarray}
 E_L^{[N]ex}
  \equiv
     E_J^{[N]}
    -E_{\mathrm{gs}}^{[N]}
  =  \frac{[2(L+\mathfrak{f)}\pm 1]L}
          {2\mathfrak{f}+1}, \label{nex}
\end{eqnarray}
where $+(-)$ is for odd (even) $N$.
As examples, the spectra of $E_L^{[N]\mathrm{ex}}$ with $\mathfrak{f}=3$ and $N=5\rightarrow9$ are shown in \fref{fig1}.
The spectra with $\mathfrak{f}=1\rightarrow 5$ and $N\geq 10$ (even) are shown in \fref{fig2}.

It turns out that $E_L^{[N]\mathrm{ex}}$ does not depend on $N$, but $L$ and $(-1)^N$, this is clearly shown in \fref{fig1}.
Say, the two spectra with $N=7$ and 9 are similar, except that the latter contains one more band at the top.
We found
\begin{eqnarray}
 E_{L+1}^{[N]\mathrm{ex}}-E_L^{[N]\mathrm{ex}}
  =  \frac{4L+2\mathfrak{f}+\mu}{2\mathfrak{f}+1}, \label{exll1}
\end{eqnarray}
where $\mu=2-(-1)^N$.
Thus, the separation between the higher bands is larger.
Recall that the excitation $L\rightarrow L+1$ is realized by breaking a [0]-pair.
Thus, the solidity of the [0]-pair does not depend on how many other [0]-pairs are surrounding, but depends on how many unpaired particles are surrounding, i.e., on the seniority of the state.
This is shown in \fref{fig1} and \fref{fig2}.
The solidity depends also on $\mathfrak{f}$ as shown in \eref{exll1}.
The [0]-pair with a larger $\mathfrak{f}$ would be easier to be broken as shown in \fref{fig2}, where the level would be lower when $\mathfrak{f}$ increases.

It is clear from the Schr\"{o}dinger equation (\ref{sch}) that, if some [0]-pairs are added into (removed from) an eigenstate so that $\Psi_{J,S,l}^{[N]}\rightarrow\Psi_{J',S,l'}^{[N']}$, where $N'-2J'=N-2J\equiv N_1$.
Then, these two states will both belong to the $L=N_1/2$ (if $N$ is even) band and have exactly the same core $\Phi_{N_1,S,l}$.
They are different only in the number of [0]-pairs.
Consequently, these states different in $N$ are related to each other.
This relationship could be called brotherhood.

Note that $M$, the $Z$-component of $S$, is a good quantum number.
We assume $M=0$ (the cases with $M\neq 0$ are similar when an external field is absent).
When $N$ is fixed, the total number of spin states with $M=0$ denoted as $\mathcal{N}_N$ is known (say, for $\mathfrak{f}=3$, $\mathcal{N}_N=8,18,32,\cdots$ when $N=3,4,5,\cdots$).
From the brotherhood, we know that, when a [0]-pair is added to every spin state of a $(N-2)$-body system, all the spin states of the $N$-body system can be recovered except those in the top-band.
Therefore, the number of states contained in the top-band is
\begin{eqnarray}
 \mathfrak{N}_{K,(-1)^N}
  =  \mathcal{N}_N
    -\mathcal{N}_{N-2}. \label{nn2}
\end{eqnarray}
In general, for the $L$ band,
\begin{eqnarray}
 \mathfrak{N}_{L,(-1)^N}
  =  \mathcal{N}_{2L}
    -\mathcal{N}_{2L-2}.
\end{eqnarray}
For the bottom band with $L=0$, $\mathfrak{N}_{0,(-1)^N}=1$.

Summing up the above numbers, we found
\begin{eqnarray}
 \sum_{L=0}^K
 \mathfrak{N}_{L,(-1)^N}
  =  \mathcal{N}_N. \label{zn}
\end{eqnarray}
\Eref{zn} confirms that the set of eigenstates $\{\Psi_{J,S,l}^{[N]}\}$ is complete.
The numbers $\mathfrak{N}_{L,(-1)^N}$ of various states are given in \fref{fig1} and \fref{fig2}.
It increases rapidly with $N_1$ (i.e., with $L$) as shown in \fref{fig1}, and with $\mathfrak{f}$ as shown in \fref{fig2}.

\section{Evaluation of the effect caused by a deviation of the Hamiltonian from singlet pairing}

\begin{figure}[htb]
 \centering \resizebox{0.95\columnwidth}{!}{\includegraphics{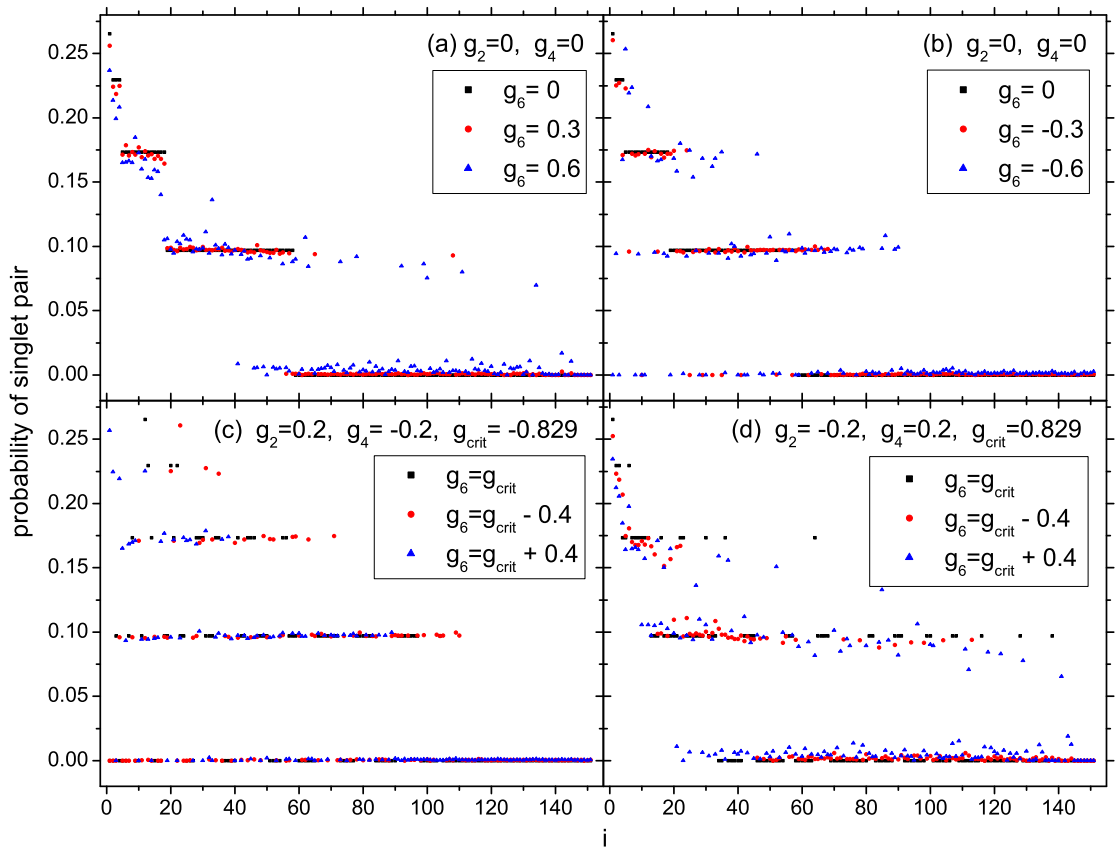} }
 \caption{
 $P_{\psi}^0$ of all the eigenstates with $N=8$, $\mathfrak{f}=3$, $M=0$, and $g_0=-1$.
 Other strengths are marked in the panels.
 $g_{24}\equiv(18g_4-11g_2)/7$ is a critical value at which the conservation of seniority holds.
 Each state is specified by an index $i$ from 1 to 151 in the order of uprising energy given as the abscissa.
 $P_{\psi}^0$ of the $i$-th state with respect to a set of strengths is marked by a point in black, red, or blue.
 }
 \label{fig3}
\end{figure}

When $H_{\mathrm{spin}}$ deviates from $H_{[0]}$, it is denoted as $H_{\mathrm{devi}}$.
In the follows we present results from exact numerical calculation to evaluate the effect of $H_{\mathrm{devi}}$.
Examples with $\mathfrak{f}=3$ and $N=8$ are given in \fref{fig3}, in which $g_0$ remains to be most negative.
Four cases are shown:
(i) $g_2=g_4\leq g_6$, in this case a shift of the strengths as a whole is further performed so that $g_2=g_4=0$ (\fref{fig3}a).
(ii) $g_2=g_4=0\geq g_6$ (\fref{fig3}b).
(iii) $g_2<g_4$ (\fref{fig3}c).
(iv) $g_2>g_4$ (\fref{fig3}d).
Let $P_{\psi}^0$ be the probability of a pair of atoms coupled to zero.
Since the key point is to check the conservation of seniority, in these figures $P_{\psi}^0$ of all the 151 eigenstates with $M=0$ are plotted.
If a group of states have the same seniority, their $P_{\psi}^0$ would be exactly the same.
Otherwise, the distribution of $P_{\psi}^0$ is diffused.

In \fref{fig3}a and \fref{fig3}b $g_2=g_4=0$ is given, while $g_6$ is given at 3 values.
In \fref{fig3}a we found that the black points (associated with $g_6=0$) are exactly distributed on five horizontal lines implying the conservation of seniority and, accordingly, the clear band structure.
The upmost line containing only one point is for the g.s. with seniority zero (all the eight particles are in [0]-pairs), while the lowest line is for the top-band with seniority being 8 (full in seniority).
When $g_6>0$ (\fref{fig3}a), we found that the points previously belonging to a horizontal line diffuse strongly along horizontal direction but weakly along vertical direction.
In particular, each diffused point has a partner point at the horizontal line, the spin states of this pair are very close to each other (see below).
It implies that all the eigenstates of $H_{[0]}$ have not been seriously disturbed by $H_{\mathrm{devi}}$, and therefore seniority remains nearly conserved.
However, their energies may be seriously affected, the order of the states (i.e., the index $i$) may therefore be changed.
E.g., when $g_6=0.3$, the lowest three states of the $L=2$ band ($i=5$, 6, and 7) have $S=3$, 0, and 8.
However, when $g_6=0.6$, the indexes of these three states become $i=5$, 9, and 6, respectively.
Due to the diffusion along horizontal direction, the band widths will become broader, and the crossover of bands may occur.
E.g., when $g_6=0.6$, the top state of the $L=2$ band has $i=33$, and its energy level goes up deeply into the $L=3$ band.
This state has $S=12$ and is found to be dominated by the component containing two [0]-pairs and four unpaired particles.
Since the four spins coupled to $S=12$ are aligned, this state has a smaller $P_{\psi}^0$ and is subjected to a stronger repulsion from $g_6$.
This explains why this state is higher.
Similarly, when $g_6=0.6$, the highest state of the $L=3$ band has $i=134$, it goes up deeply into the $L=4$ band.
This state has $S=18$ and is dominated by one [0]-pairs and six unpaired particles.
Since the six spins coupled to $S=18$ are all aligned, this state has also a smaller $P_{\psi}^0$ and also subjected to a stronger repulsion from $g_6$.
Whereas we found that the lowest state of the top-band has $i=41$ and $S=0$.
Therefore, the repulsion from $g_6$ can be reduced.
Thus, although the seniority is nearly conserved, the energies are strongly affected by $H_{\mathrm{devi}}$.

To evaluate the deviation quantitatively, let the eigenstates of $H_{[0]}$ be denoted as $\Phi_{i'}$, while those of $H_{\mathrm{devi}}$ be $\Psi_i$.
For the case given in \fref{fig3}a with $g_6=0.3$, among the 151 eigenstates, 81 of them can find a partner so that $\langle\Phi_{i'}|\Psi_i\rangle>0.99$ and other 61 of them have $0.95<\langle\Phi_{i'}|\Psi_i\rangle<0.99$.
These data confirm that the deviation in the eigenstates as a whole is slight.
In particular, the deviation in high-lying states is even much smaller.
Say, $\langle\Phi_{i'}|\Psi_1\rangle=0.950$, $\langle\Phi_{i'}|\Psi_{21}\rangle=0.993$, and $\langle\Phi_{i'}|\Psi_{150}\rangle=1.000$.
When $g_6$ increases from $0.3$ to $0.6$, the above value $81\rightarrow 22$ and $61\rightarrow 60$, it implies a larger deviation.

In \fref{fig3}b the vertical diffusion of points is also slight, thus the near conservation of seniority also holds, in particular for higher states.
However, the widths of the bands become very broad.
When $g_6=-0.6$, the energy of the bottom state of every band is close to the g.s..
For examples, the bottom state of the top-band with $L=4$ has $i=1$ and $S=24$.
Thus, this state is fully polarized and it becomes the g.s..
Accordingly, the width of the top-band covers all the spectrum.
When both $g_6$ and $g_0$ are negative, there is a competition between alignment and pairing.
When alignment exceeds pairing, as in the above case, a change of phase of the g.s. would occur.
Similarly, the blue point for the bottom state of the $L=3$ band has $i=2$ and $S=18$.
It is composed of 6 aligned particles together with a [0]-pair.
While the blue point for the bottom state of the $L=2$ band has $i=4$ and $S=12$ containing four aligned particles and two [0]-pairs.
Thus, when $g_6$ is sufficiently negative, all the unpaired particles are aligned in the bottom states of every band.
Alternatively, the attraction from $g_6$ would be minimized in $S=0$ states.
Therefore, among the five bands, four top states have $S=0$ (the appearance of five $S=0$ states is prohibited because, for $N=8$ and $\mathfrak{f}=3$, the multiplicity of $S=0$ state is 4).

For $\mathfrak{f}=3$, it has been pointed out that there is a critical value $g_6=(18g_4-11g_2)/7\equiv g_{\mathrm{crit}}$ (disregarding the value of $g_0$) at which the seniority is strictly conserved \cite{pvi}.
\Fref{fig3}c and \fref{fig3}d are for the case in the neighborhood of $g_{\mathrm{crit}}$.
In \fref{fig3}c, $g_2>g_4$, and $g_{\mathrm{crit}}$ becomes negative.
Accordingly, due to the attraction from $g_6$, this figure is similar to \fref{fig3}b.
Whereas in \fref{fig3}d, $g_2<g_4$, and $g_{\mathrm{crit}}$ becomes positive.
Accordingly, due to the repulsion from $g_6$, this figure is more or less similar to \fref{fig3}a.
Note that, for \fref{fig3}a, \fref{fig3}b, and when $g_6=0$, all the states belonging to the same band are degenerate.
However, when $g_6=g_{\mathrm{crit}}$, the states belonging to the same band are not degenerate.
The band is further divided into a few pieces.
Only the states in a piece are degenerate \cite{pvi}.

\begin{figure}[htb]
 \centering \resizebox{0.95\columnwidth}{!}{\includegraphics{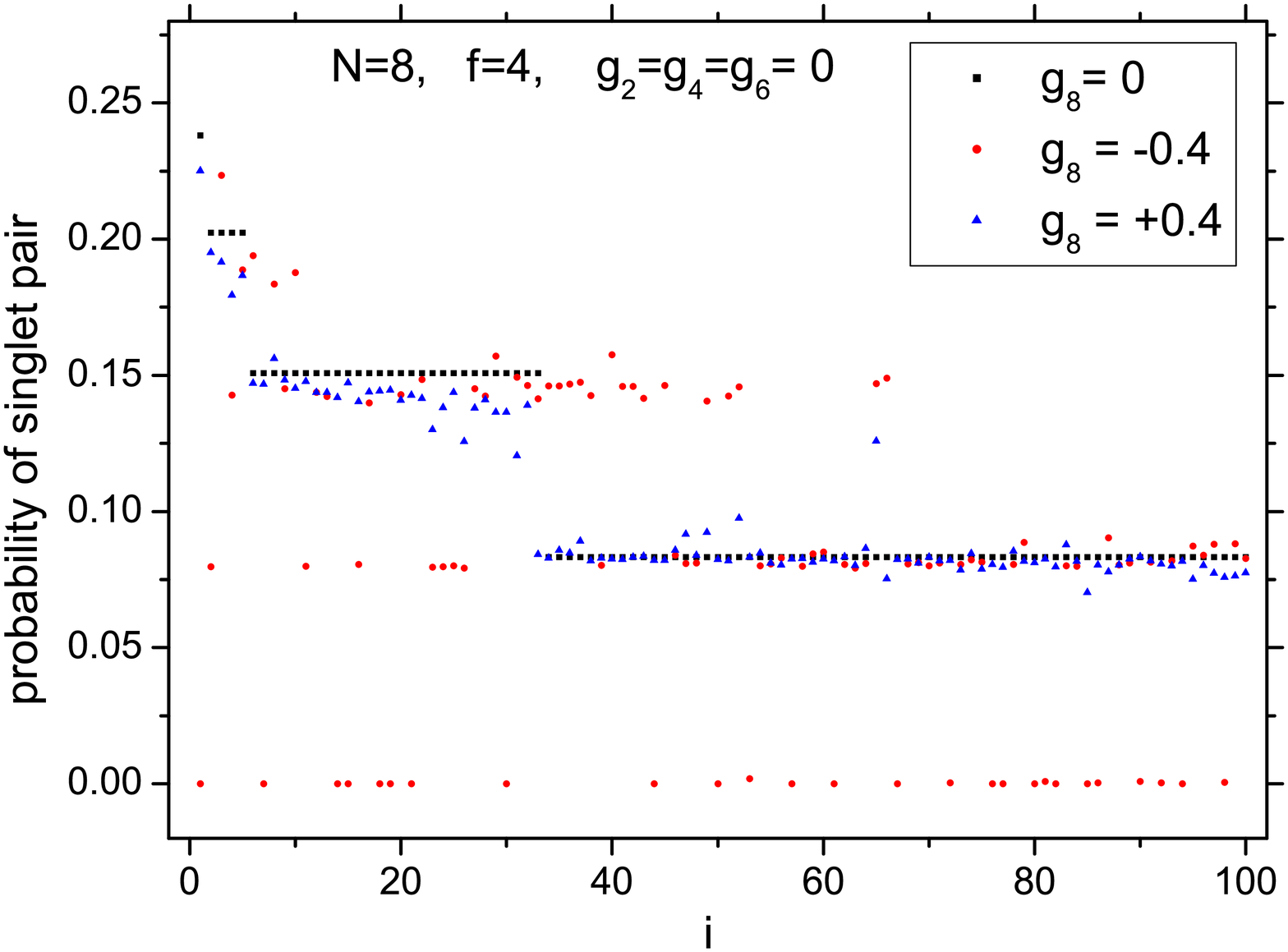} }
 \caption{
 Refer to \fref{fig3} but with $\mathfrak{f}=4$.
 }
 \label{fig4}
\end{figure}

One more example with $\mathfrak{f}=4$ is given in \fref{fig4}.
For $N=8$, there are totally 526 $M=0$ eigenstates.
$P_{\psi}^0$ of the lowest 100 states are plotted.
For higher states with $i>100$, $P_{\psi}^0$ of them are $<0.09$.
When $i>250$, most $P_{\psi}^0$ are close to zero.
\Fref{fig4} demonstrates that, with $H_{\mathrm{devi}}$, a larger $\mathfrak{f}$ does not further spoil the near conservation of seniority.
The corresponding band structure remains clear.

\section{Final remarks}

For $H_{[0]}$, we have solved the $N$-boson problem with arbitrary spin $\mathfrak{f}>0$.
The complete spectra for systems different in $N$ and $\mathfrak{f}$ have been obtained analytically.
The completeness of the spectra has been proved.
It is found that the band structure based on the conservation of seniority holds also when $\mathfrak{f}\geq 3$.
Comparison has been made among the spectra different in $N$ and/or $\mathfrak{f}$.
Similarity and relationship among their eigenstates have been demonstrated.
In particular, for $H_{\mathrm{devi}}$, the effect of the deviation has been studied via strict calculations on $P_{\psi}^0$.
The following points are mentioned.

\begin{itemize}
\item
There is brotherhood among the eigenstates different in $N$.
Therefore, the knowledge from few-body systems (which can be obtained rigorously) can be used to evaluate the low-lying states of many-body systems.

\item
The case with a positive $g_0$ is noticeable.
When all the strengths change their signs ($g_0\rightarrow+1$), all the above discussions remain valid unless that the whole spectrum is reversed.
In this case, the ground band might have an extremely high level density and well protected by a very large energy gap $E_{\mathrm{gap}}=(2N+2\mathfrak{f}-3)/(2\mathfrak{f}+1)$.
Let $N=1000$ as an example.
When $\mathfrak{f}=2$, the number of $M=0$ states contained in the ground band is $1.672\times 10^5$.
However, when $\mathfrak{f}=3$, this number is $7.716\times 10^9$.
The great difference demonstrates that the increase of $\mathfrak{f}$ could lead to extremely high level density.
If the strengths could be tuned so that $g_2$ is close to $g_4<g_0$ (for $\mathfrak{f}=2$) or $g_2$ and $g_4$ are both close to $g_6<g_0$ (for $\mathfrak{f}=3$), the level density of the ground band could be extremely large.

\item
For $\mathfrak{f}=3$, there are two zones in the parameter space lying along the line $g_2=g_4=g_6$ and the line $g_6=(18g_4-11g_2)/7$ in which the conservation of seniority holds roughly.
In these zones the whole spectra can be divided into bands.
The states in a band have their $P_{\psi}^0$ close to each other.
For $\mathfrak{f}\geq 4$, it is believed that there would also be zones in which the band structure holds.
Nonetheless, the case with $\mathfrak{f}\geq 4$ remains to be further clarified.
\end{itemize}

\ack
This work is supported by the National Natural Science Foundation of China under Grants No.11874432, 11372122, 10874122, and by the Key Area Research and Development Program of Guangdong Province under Grant No. 2019B030330001.

\end{document}